\documentstyle[12pt]{article}
\newcommand{\be}{\begin{equation}}
\newcommand{\ee}{\end{equation}}
 
\newcommand{\bea}{\begin{eqnarray}}
\newcommand{\eea}{\end{eqnarray}}
\newcommand{\ba}{\begin{array}}
\newcommand{\ea}{\end{array}}
\newcommand{\beqa}{\begin{eqnarray}}
\newcommand{\eeqa}{\end{eqnarray}}

\newcommand{\NP}[1]{Nucl. Phys.\ {\bf #1}}
\newcommand{\PL}[1]{Phys. Lett.\ {\bf #1}}

\newcommand{\PRD}[1]{Phys. Rev.\ {\bf #1}}
\newcommand{\PR}[1]{Phys. Rep.\ {\bf #1}}
\newcommand{\PRL}[1]{Phys. Rev. Lett.\ {\bf #1}}

\newcommand{\ZP}[1]{Z. Phys.\ {\bf #1}}

\newcommand{\Tr}{{\rm Tr}}

\newcommand{\ssu}{$SU(2)_L\times SU(2)_R\times U(1)_{B-L}\,$}

\newcommand{\matr}{\left( \begin{array}}
\newcommand{\ematr}{\end{array} \right)}

\newcommand{\lsim}
{{\;\raise0.3ex\hbox{$<$\kern-0.75em\raise-1.1ex\hbox{$\sim$}}\;}}
\newcommand{\gsim}
{{\;\raise0.3ex\hbox{$>$\kern-0.75em\raise-1.1ex\hbox{$\sim$}}\;}}

\begin{document}

\begin{titlepage}

\begin{flushright}
{\large HIP-1997-06/TH}\\
\end{flushright} 
 
\Large
 
\begin{center}
{\bf Constraining the vacuum expectation values in naturally
R-parity conserving supersymmetric models}
 
\bigskip
\bigskip

\normalsize
{K. Huitu$^a$, P.N. Pandita$^{a,b}$
and K. Puolam\"aki$^a$}\\
[15pt]
{\it $^a$Helsinki Institute of Physics, FIN-00014 University of Helsinki,
Finland
\\$^b$Department of Physics, North Eastern Hill University,
Shillong 793022, India\footnote{Permanent address} }

\vspace{2.truecm}

{\bf\normalsize \bf Abstract}

 \end{center}

{\normalsize

We obtain a relation  between right-handed gauge boson mass and soft 
squark mass in naturally R-parity conserving general supersymmetric left-right 
models.
This relation implies that either ${W_R}$ is lighter than twice the
soft squark mass, or a ratio of vacuum 
expectation values (VEVs) in the model, denoted by $\tan\alpha$, is close 
to its value of unity in 
the limit of vanishing $D$-terms.
Generally, we find that for heavy $W_R$ $\tan\alpha$ is larger 
than one, and that the
right-handed sneutrino VEV, responsible for spontaneous $R$-parity
breaking, is at most of the order $M_{SUSY}/h_{\Delta_R}$, where
$M_{SUSY}$ is supersymmetry breaking scale and $h_{\Delta_R}$ is the
Yukawa coupling in Majorana mass term for right-handed neutrinos.
These constraints follow from $SU(3)_c$ and $U(1)_{em}$ gauge invariance of 
the ground state of the theory.

\noindent\begin{flushleft}
\end{flushleft}}

\vfill

\normalsize

\end{titlepage}
 
\newpage
 
\setcounter{page}{2}

Low energy supersymmetry is at present the only known extension of the 
Standard 
Model (SM) in which elementary Higgs scalar field, which is necessary to break
the $SU(2)_L\times U(1)_Y$ symmetry, is natural \cite{susy}.
The minimal supersymmetric standard model (MSSM) is constructed by
simply doubling the number of degrees of freedom of the Standard Model, and 
adding an extra
Higgs doublet (with opposite hypercharge) to cancel gauge anomalies and to 
generate masses for all quarks and leptons.
However, one of the successes of SM, namely the automatic conservation of
baryon number (B) and lepton number (L) by the renormalizable
interactions, is not shared by the minimal supersymmetric standard model.
In SM, the conservation of B and L follows from the particle content 
and the $SU(3)_c\times SU(2)_L\times U(1)_Y$ gauge invariance.
In MSSM, baryon and lepton
number violation can occur at tree level with catastrophic consequences unless
the corresponding couplings are very small.
The most common way to eliminate these tree level B and L violating terms is 
to impose a discrete $Z_2$ symmetry \cite{BL1,BL2} known as a matter parity
for superfields (=$(-1)^{3(B-L)}$) or equivalently
R-parity on component fields ($R_p=(-1)^{3(B-L)+2S}$, $S$ being the spin
of the particle), with all the Standard Model particles having 
$R_p=+1$, while all the superpartners have $R_p=-1$.
However, the assumption of R-parity conservation appears to be {\it ad hoc},
since it is not required for the internal consistency of the minimal
supersymmetric standard model.
Furthermore, all global symmetries, discrete or continuous, could be violated
by the Planck scale physics effects \cite{planck}.
The net effect would be
the appearance of extremely tiny violations of otherwise exact global
symmetries in the Lagrangians describing physics at the electroweak scale.
The problem becomes acute for low energy supersymmetric models \cite{Rpar} 
because
B and L are no longer automatic symmetries of the Lagrangian as they are 
in the Standard Model.

It would, therefore, be more appealing to have a supersymmetric theory
where R-parity is related to a gauge symmetry, and its conservation
is automatic because of the invariance of the underlying theory under an 
extended gauge symmetry.
Indeed $R_p$ conservation follows automatically in certain theories with
gauged $(B-L)$, as is suggested by the fact that matter parity is simply a
$Z_2$ subgroup of $(B-L)$.
It has been noted by several authors \cite{Rauto1,Rauto2} that if the gauge 
symmetry 
of MSSM is extended to $SU(2)_L\times U(1)_{I_{3R}}\times U(1)_{B-L}$ or
$SU(2)_L\times SU(2)_R\times U(1)_{B-L}$, the theory becomes
automatically R-parity conserving.
Such a left-right supersymmetric theory (SUSYLR) solves the problems of
explicit B and L violation of MSSM and has received much attention
recently $\left[\right.$8-12$\left.\right]$.
In particular, the dynamical issues connected with automatic R-parity
conservation have been considered in case of left-right supersymmetric
models \cite{km}.
It has been found in a wide class of such models \cite{km} that R-parity must 
be 
spontaneously broken \cite{am} because of the form of the scalar potential,
although this could be avoided if non-renormalizable interactions are 
included.
Thus, this model cures one of the major drawbacks of MSSM, although it leads
to small L-violating terms, which are, however, suppressed and could be
tested experimentally by searching for lepton number violation.

Furthermore, it has been shown \cite{km2} in the minimal SUSYLR model
that the mass ($m_{W_R}$) of the right-handed gauge boson $W_R$ has an upper
limit related to the SUSY breaking scale, {\it i.e.}, 
$m_{W_R}\leq g M_{SUSY}/h_{\Delta_R}$, where $g$ is the weak gauge coupling and
$h_{\Delta_R}$ is the Yukawa coupling of the right-handed neutrinos 
with the triplet Higgs fields.
This result is a consequence of electric charge conservation and low
energy parity violation by the ground state of the theory.

In this letter we report on further consequences of SUSYLR
models.
For phenomenological reasons, it is desirable to constrain the
magnitudes of the $SU(2)_R$ breaking scale, represented by 
$m_{W_R}$, and R-parity
breaking scale, corresponding to the right-handed sneutrino VEV.
Here we will find a relation between the $W_R$ mass and the soft 
squark mass $\tilde m$, independently of any Yukawa couplings.  
In addition, we constrain the order of the right-handed sneutrino VEV to 
be generally at most $M_{SUSY}/h_{\Delta_R}$ for large $W_R$ mass.
These results follow from the conservation of electric charge and color 
by the ground state of the theory.

We begin by recalling the basic features of the SUSYLR models.
The matter fields of the minimal model consist of the quark and lepton
doublets, $Q(2,1,1/3)$; $Q^c(1,2,-1/3);\; L(2,1,-1);\; L^c(1,2,1)$,
and the Higgs multiplets consist of $\Delta_L(3,1,-2)$; $\Delta_R(1,3,-2);
\; \delta_L(3,1,2);\;\delta_R(1,3,2);\; \Phi (2,2,0);\; \chi (2,2,0)$.
The numbers in the parantheses denote the representation content of the
fields under the gauge group \ssu.
The most general superpotential of the model is given by 

\bea
W&=& h_{\phi Q}Q^T i\tau_2 \Phi Q^c +h_{\chi Q}Q^T i\tau_2 \chi Q^c +
h_{\phi L}L^T i\tau_2 \Phi L^c 
+ h_{\chi L}L^T i\tau_2 \chi L^c \nonumber\\
&&+h_{\delta_L} L^T i\tau_2 \delta_L L +
h_{\Delta_R} L^{cT} i\tau_2 \Delta_R L^c+
\mu_1 \Tr (i\tau_2\Phi^T i\tau_2 \chi) +
\mu_1' \Tr (i\tau_2\Phi^T i\tau_2 \Phi) \nonumber\\
&&+ \mu_1'' \Tr (i\tau_2\chi^T i\tau_2 \chi) 
+\Tr (\mu_{2L}\Delta_L \delta_L +
\mu_{2R}\Delta_R\delta_R).
\eea

The general form of the Higgs potential is given by

\be
V=V_F+V_D+V_{soft}
\label{pot}
\ee

\noindent
and can be calculated in a straightforward manner.
In the following we shall represent the scalar components of the superfields
by the same symbols as the superfields themselves.
The most general form of the vacuum expectation values of various scalar fields
which preserves $Q_{em}$ can be written as

\bea
&&\langle \Phi\rangle  = \matr {cc} \kappa_1&0\\0&e^{i\varphi_1}\kappa '_1 \ematr 
,\;\;
\langle \chi\rangle  = \matr {cc} e^{i\varphi_2}\kappa '_2&0\\0&\kappa_2 \ematr ,
\nonumber\\
&& \langle \Delta_L\rangle = \matr {cc} 0&v_{\Delta_L}\\0&0\ematr ,\;
\langle \delta_L\rangle = \matr {cc} 0&0\\v_{\delta_L}&0\ematr , \nonumber\\
&&\; \langle \Delta_R\rangle = \matr {cc} 0&v_{\Delta_R}\\0&0\ematr ,\;
\langle \delta_R\rangle = \matr {cc} 0&0\\v_{\delta_R}&0\ematr ,\nonumber\\
&&\langle L\rangle =\matr {c} \sigma_L\\0\ematr,\;
\langle L^c\rangle =\matr {c} 0\\\sigma_R\ematr .
\eea

For simplicity, we shall ignore the phases $\varphi_1$ and $\varphi_2$
in the following, although this does not affect the final conclusion.
Due to the tiny mixing between the charged gauge bosons, $\kappa '_1$
and $\kappa '_2$ are taken to be much smaller than $\kappa_1$ and $\kappa_2$.
Furthermore, since the electroweak $\rho$-parameter is close to unity,
$\rho=1.0002\pm0.0013\pm0.0018 $ \cite{pdg}, the triplet vacuum expectation
values $\langle\Delta_L\rangle $ and $\langle\delta_L\rangle $ must be small.
For definiteness, we shall take 
$v_{\Delta_R}\sim v_{\delta_R}\sim v_R$, the generic scale of the
right-handed symmetry breaking.
In this class of models the spontaneous breakdown of R-parity is inevitable 
\cite{km}, and therefore we shall assume that $\sigma_L$ and $\sigma_R$
are non-zero.
Since $\sigma_L$ contributes to $W_L$ mass, it is less than the weak scale.
On the other hand, in electric charge preserving ground state
$\sigma_R$ is necessarily at least of the
order of the typical
SUSY breaking scale $M_{SUSY}$  or the right-handed breaking scale
$v_R$,
whichever is lower \cite{km2}.

We next construct the up- and
down-squark mass matrices.
To this end we write down explicitly the different components of the
scalar potential (\ref{pot}) ($g_L,\; g_R,\; g_{B-L}$ are the gauge couplings)

\bea
V_F&=&|h_{\phi Q}Q^cQ^T(i\tau_2)+h_{\phi L}L^cL^T(i\tau_2)+
\mu_1(i\tau_2)\chi^T(i\tau_2) +2\mu'_1(i\tau_2)\Phi^T(i\tau_2)|^2\nonumber\\
&&+
|h_{\chi Q}Q^cQ^T(i\tau_2)+h_{\chi L}L^cL^T(i\tau_2)+
\mu_1(i\tau_2)\Phi^T(i\tau_2) + 2\mu''_1(i\tau_2)\chi^T(i\tau_2)|^2\nonumber\\
&& + |(i\tau_2)(h_{\phi Q}\Phi +h_{\chi Q}\chi )Q^c|^2+
|Q^T(i\tau_2)(h_{\phi Q}\Phi +h_{\chi Q}\chi )|^2,
\label{Fterm}
\eea

\bea
V_D&=& \frac 18 g_L^2\sum_a (\Tr (\Phi^\dagger\tau_a\Phi )+
\Tr (\chi^\dagger\tau_a\chi ) 
+2\Tr (\Delta_L^\dagger\tau_a\Delta_L )
+2\Tr (\delta_L^\dagger\tau_a\delta_L )
\nonumber\\
&&+L^\dagger\tau_a L+Q^\dagger\tau_a Q )^2
+\frac 18 g_R^2\sum_a (-\Tr (\Phi\tau_a\Phi^\dagger )-
\Tr (\chi\tau_a\chi^\dagger ) \nonumber\\
&&+2\Tr (\Delta_R^\dagger\tau_a\Delta_R )
+2\Tr (\delta_R^\dagger\tau_a\delta_R )
+L^{c\dagger}\tau_a L^c+Q^{c\dagger}\tau_a Q^c )^2
\nonumber\\
&&+\frac 18 g_{B-L}^2(-2\Tr (\Delta_R^\dagger\Delta_R)+
2\Tr (\delta_R^\dagger\delta_R)-2\Tr (\Delta_L^\dagger\Delta_L)+
2\Tr (\delta_L^\dagger\delta_L)
\nonumber\\ &&
-L^\dagger L+L^{c\dagger} L^c
+\frac 13 Q^\dagger Q-\frac 13 Q^{c\dagger} Q^c )^2,
\label{Dterm}
\eea

\bea
V_{soft}&=&\tilde m_Q^2|Q|^2+\tilde m_{Q^c}^2|Q^c|^2+
(Q^Ti\tau_2(B_\phi \Phi +B_\chi \chi )Q^c +h.c.),
\label{softterm}
\eea

\noindent
where we have retained only those terms which are relevant for our
discussion.

{}From (\ref{Fterm}), (\ref{Dterm}), and (\ref{softterm})
it is straightforward to construct the various squark mass matrices.
For definiteness we consider the up- and down-squark mass matrices for the 
lightest generation (ignoring the intergenerational mixing) which gives
the tightest constraint in our case.
The part of the potential containing the squark mass terms can be
written as

\be
V_{squark}=\matr {cc}   U_L^* &   U_R^* \ematr \tilde M_U
\matr {c}   U_L \\   U_R \ematr +
\matr {cc}  D_L^* &   D_R^* \ematr \tilde M_D
\matr {c}   D_L \\   D_R \ematr .
\ee

\noindent 
The mass matrix elements for the up-type squarks are

\bea
(\tilde M_U)_{  U_L^*   U_L} &=&  \tilde m_Q^2+
m_u^2 +
\frac 14 g_L^2(\omega^2_\kappa -2\omega^2_L ) +
\frac 16 g_{B-L}^2(\omega^2_L -\omega^2_R ) ,
\nonumber\\
(\tilde M_U)_{  U_R^*   U_L} &=&
B_\phi \kappa '_1+B_\chi\kappa_2 - 
\mu_1 (h_{\phi Q}\kappa '_2+h_{\chi Q}\kappa_1)-
2h_{\phi Q}\mu '_1 \kappa_1-
2 h_{\chi Q}\mu''_1\kappa '_2\nonumber\\
&&+
(h_{\phi L }h_{\phi Q } +h_{\chi L }h_{\chi Q }) \sigma_L\sigma_R \nonumber\\
&=&\left[(\tilde M_U)_{  U_L^* U_R }\right]^*,\nonumber\\
(\tilde M_U)_{  U_R^*   U_R} &=&\tilde m_{Q^c}^2+
m_u^2 +
\frac 14 g_R^2(\omega^2_\kappa -2\omega^2_R ) +
\frac 16 g_{B-L}^2(\omega^2_R -\omega^2_L ) 
,
\eea

\noindent
and for down-type squarks

\bea
(\tilde M_D)_{  D_L^*   D_L} &=&
 \tilde m_Q^2+
m_d^2
-\frac 14 g_L^2(\omega^2_\kappa -2\omega^2_L ) +
\frac 16 g_{B-L}^2(\omega^2_L -\omega^2_R ) 
,\nonumber\\
(\tilde M_D)_{  D_R^*   D_L} &=&
-B_\phi \kappa_1-B_\chi\kappa '_2 + 
\mu_1 (h_{\phi Q}\kappa_2+h_{\chi Q}\kappa '_1) +
2h_{\phi Q}\mu '_1 \kappa '_1+
2 h_{\chi Q}\mu''_1\kappa_2
\nonumber\\
& = &\left[(\tilde M_D)_{  D_L^* D_R  }\right]^*,\nonumber\\
(\tilde M_D)_{ D_R^*  D_R  } &=&\tilde m_{Q^c}^2+
m_d^2
-\frac 14 g_R^2(\omega^2_\kappa -2\omega^2_R ) +
\frac 16 g_{B-L}^2(\omega^2_R -\omega^2_L ) ,
\eea

\noindent
where

\be
m_u=h_{\phi Q}\kappa '_1+h_{\chi Q}\kappa_2,\;
m_d=h_{\phi Q}\kappa_1+h_{\chi Q}\kappa '_2.
\ee

\noindent
and

\bea
\omega^2_L=v_{\delta_L}^2-v_{\Delta_L}^2-\frac 12\sigma_L^2,\;
\omega^2_R=v_{\Delta_R}^2-v_{\delta_R}^2-\frac 12\sigma_R^2,\;
\omega^2_\kappa = \kappa_1^2+{\kappa '_2}^2-\kappa_2^2-{\kappa '_1}^2.
\eea

In order not to break electromagnetism or color, none of the physical
squared masses of squarks can be negative.
Necessarily then all the diagonal elements of the squark mass matrices
should be non-negative.
Since the SUSY breaking scale is expected to be ${\cal{O}}(1$ TeV),
the weak scale is smaller than the other scales involved, $v_R$,
$\sigma_R$ or $M_{SUSY}$.
We'll ignore the terms of the order of the weak scale or smaller
in the following.
Combining the diagonal elements of the mass matrices $\tilde M_U $ and
$\tilde M_D$, it follows that

\be
\tilde m_Q^2+\tilde m_{Q^c}^2 \geq |\frac 12 g_R^2\omega^2_R|=
\frac 12g_R^2|v_{\Delta_R}^2-
v_{\delta_R}^2 -\frac 12 \sigma_R^2 |.
\label{wr1}
\ee

\noindent
In arriving at (\ref{wr1}), we have made no assumptions other than
that of ignoring those quantities which are of the order of weak scale.

We define next an angle $\alpha $ with 
$\tan^2\alpha =(v_{\delta_R}^2 +\frac 12 \sigma_R^2)/ v_{\Delta_R}^2$ 
and write
$\tilde m_Q^2=\tilde m_{Q^c}^2\equiv\tilde m^2$.
Then Eq.(\ref{wr1}) can be written as 
\bea
\tilde m^2\geq \frac 14 g_R^2v_{\Delta_R}^2 |1-\tan^{2}\alpha|.
\label{wr2}
\eea

In order to understand the result (\ref{wr1}) or (\ref{wr2}), we recall that 
the right-handed
gauge boson mass is given by (ignoring weak scale effects) \cite{hm}

\be
m_{W_R}^2=g_R^2 (v_{\Delta_R}^2+v_{\delta_R}^2 +\frac 12\sigma_R^2 )
=g_R^2 v_{\Delta_R}^2 (1+\tan^{2}\alpha ).
\label{mwr}
\ee

\noindent
Combining (\ref{wr2}) and  (\ref{mwr}), we find

\be
m_{W_R}^2|\cos 2\alpha|\leq 4\tilde m^2.
\label{result}
\ee

If the $W_R$ boson is lighter than twice the soft squark mass $\tilde m$,
Eq.(\ref{result}) 
is fulfilled for any $\tan\alpha $.
If $\tilde m\sim M_{SUSY}\sim 1$ TeV as is commonly assumed, 
$m_{W_R}$ cannot be 
much less, since experimentally  $m_{W_R}>420$ GeV \cite{cdf}.
On the other hand,
if $m_{W_R}$ is much larger than $2\tilde m$, $\tan\alpha$ has to be close
to one, e.g. for $m_{W_R}=10$ TeV and $\tilde m$=1 TeV, one would need
$0.96\leq\tan\alpha\leq 1.04$.
It is interesting to note in this context that 
the vanishing of $D$-terms implies $\tan\alpha=1$.
To translate the limit for $\tan\alpha $ to  an upper bound for the 
VEV $\langle \Delta_R^0\rangle$, one needs a lower limit for
$g_R$.
This was found in \cite{cl} from 
$\sin^2\theta_W=e^2/g_L^2=0.23$, namely $g_R \geq 0.55\; g_L$.
Consequently 
$v_{\Delta_R}\leq m_{W_R} |\cos\,\alpha |/(0.55 \, g_L)$, e.g.
in our example $v_{\Delta_R}\lsim 20$ TeV.

To further analyze the situation with large $m_{W_R}$,
we note that, if the right-handed scale and R-parity breaking scale differ
from each other, one has  $v_{\Delta_R},\; v_{\delta_R}
>\sigma_R$, since $\tan\alpha\sim 1$.
We recall then the doubly charged Higgs mass matrix \cite{hm} 
 given by
(ignoring terms suppressed by $\sigma_R/v_{\Delta_R}$ or
$\sigma_R/v_{\delta_R}$) 
\bea
&&M_{\Delta^{++}\delta^{++}}^2=
\matr{cc} m_{\Delta\delta}^2\frac{v_\delta}{v_\Delta}
-4h_\Delta^2\sigma_R^2-2g_R^2\omega_R^2
&-m_{\Delta\delta}^2 \\ -m_{\Delta\delta}^2 &
m_{\Delta\delta}^2\frac{v_\Delta}{v_\delta}+
2g_R^2\omega_R^2 \ematr,
\eea
where $ m_{\Delta\delta}$ is the soft parameter mixing right-handed
Higgs triplets.
The two eigenvalues of the mass matrix need to be real and non-negative
in order not to break $U(1)_{em}$.
This leads to two conditions:
\bea
&& h_{\Delta_R}^2\sigma_R^2 \leq \frac 14
m_{\Delta\delta}^2\left( \frac {v_{\delta_R}}{v_{\Delta_R}} +
 \frac {v_{\Delta_R}}{v_{\delta_R}}\right)
\label{cond1}
\eea
and
\bea
 -2m_{\Delta\delta}^2g_R^2\omega_R^2
\left( \frac {v_{\delta_R}}{v_{\Delta_R}} -
 \frac {v_{\Delta_R}}{v_{\delta_R}}\right)+4 g_R^4\omega_R^4 
+4 m_{\Delta\delta}^2h_{\Delta_R}^2\sigma_R^2\frac {v_{\Delta_R}}{v_{\delta_R}}
+ 8 h_{\Delta_R}^2\sigma_R^2 g_R^2\omega_R^2 
\leq 0
\label{cond2}
\eea

{}From (\ref{cond1}) we see that $h_{\Delta_R}\sigma_R$ can be at most of the
order of $m_{\Delta\delta}\sim M_{SUSY}$.
To fulfill the inequality  (\ref{cond2}) we can consider two
cases: $v_{\Delta_R}<v_{\delta_R}$ and  $v_{\delta_R}<v_{\Delta_R}$.
In both cases one must have $\omega_R^2\leq 0$ or
equivalently $\tan\alpha \geq 1$.
The equality can hold for $\sigma_R=0$ and $v_{\delta_R}=v_{\Delta_R}$.

In conclusion, we have studied the implications of SUSY 
left-right models, which
naturally incorporate R-parity conservation.
We have found that the $W_R$ mass and the soft squark mass are related
by (\ref{result}), which implies that either
the scale of the right-handed gauge symmetry breaking must
be close to the SUSY breaking scale,
or $\tan\alpha \sim 1$ corresponding to vanishing $D$-terms.
In general, we have also found that for large $m_{W_R}$, 
the right-handed sneutrino VEV is constrained to be at most of the
order $M_{SUSY}/h_{\Delta_R}$, and that $\tan\alpha $ is larger than one.

\noindent
{\bf Acknowledgements}

One of us (PNP) would like to thank the Helsinki Institute of Physics for 
hospitality while this work was completed.
The work of PNP is supported by the Department of Atomic Energy Project
No.37/14/95-R \& D-II/663.

\end{document}